\documentclass[11pt]{article}
\usepackage[left=2cm,top=2cm,right=2cm]{geometry}

\usepackage{hyperref} 

\usepackage[pdftex]{graphicx}
\usepackage[font=small,format=plain,labelfont=bf,up,textfont=it,up]{caption}
\usepackage{chngcntr}

\counterwithin{figure}{section}
\counterwithin{equation}{section}

\newcommand{\bs}{\bigskip}

\newcommand{\bc}{\begin{center}}
\newcommand{\ec}{\end{center}}

\begin{document}
\bs
\bc 
{\Large\bf{Theoretical Review on QCD and Vector Mesons} \\ {in Dileptonic Quark Gluon Plasma.}}
\bs

\begin{center}
\large{Vlasios Petousis} ~$^{a)}$  
\footnote{petousis@ucy.ac.cy}
\bs

$^{a)}$\small{Department of Physics, University of Cyprus,  Nicosia, CY-1678, Cyprus}

\ec

\small\date\center{\today}
\end{center}

\begin{abstract}
After the discovery of the Quantum Chromodynamics (QCD), many experimental and theoretical efforts have been made to investigate physics issues involved in ultra relativistic heavy-ion collisions. The fundamental goal of this work is to present a short guide of the underlying theory of strong interactions with emphasis on the light vector mesons based on the exceptional work of R.Rapp and J.Wambach \cite{30}. Today widely believed that deconfinement and chiral symmetry restoration probably takes place in a hot and dense hadronic medium as this produced at these collisions. We focus at the place where the examination of the new form of matter, so-called Quark Gluon Plasma (QGP), can be done through the investigation of the di-leptonic production coming from light vector mesons. Significant progress has been made since the last years, but still some open questions remain at the non-perturbative mass region.
\end{abstract}

\newpage
\tableofcontents
\newpage

\section{Introduction} 
Almost the last fifteen years a lot of experimental and theoretical efforts have been made to investigate the physics behind of  the  Ultra Relativistic Heavy Ion Collisions (URHIC's). The principal aim of this article divided in two parts. First is in general an effort to explore the structure of the underlying theory of strong interactions in Quantum Chromodynamics (QCD). Secong is to review the theoretical tools which are necessary of understanding the experimental results at the low invariant mass region, which is intimately connected to the question of chiral symmetry restoration.

\subsection{Ultra Relativistic Heavy Ion Collisions, a quick review}
The primary objective of particle physics is to discover the fundamental forces and symmetries, and the elementary particles in Nature. A hierarchy of constituents of matter has been observed: macroscopic matter consists of molecules and atoms, the atoms consist of nucleons which in turn are formed of quarks, antiquarks and gluons (partons). These results have been obtained by scattering experiments at higher and higher energies, as required to achieve information on smaller and smaller objects. 
At the moment the hierarchy ends at quarks: no substructure has been observed for them, so they are regarded as point like particles. Isolated single free quarks have never been observed, and therefore it is conjectured that quarks are confined together with other quarks to form hadrons. The strong (color) force field between the quarks is intermediated by gluons, and inside the hadrons quark-antiquark pairs form as quantum fluctuations. The fundamental theory describing the mutual strong interactions of quarks, gluons and antiquarks is Quantum Chromodynamics (QCD). 

Quantum Chromodynamics is the SU(3) gauge symmetric part of the Standard Model of particle physics. In the highest-energy lepton-proton and proton-antiproton collisions, the individual scattering processes of quarks and gluons can be directly observed in certain special cases, such as jet production, where the momentum transfer is large and the interaction takes place within a small distance compared with the size of the hadron. These processes can be treated by means of perturbation theory of QCD, i.e. their cross sections can be formulated as an expansion in powers of the strong coupling α(Q) as long as the coupling stays clearly smaller than unity. This is the case when the typical momentum (or energy) scale Q involved in the process is clearly larger than the inverse size of the hadron, i.e. when Q is much larger than the inherent momentum scale of QCD, where ${{\Lambda_{QCD}} \sim {200 MeV}}$.

In an ideal experiment of high energy particle physics the individual scattered quarks and gluons fly practically freely away, dress with a gluon cloud and rapidly form color singlet bound states, hadrons. The situation radically changes, however, if sufficiently many partons are made to scatter simultaneously into the same volume element: a dense medium of partons is formed, where the interactions of quarks, antiquarks and gluons are so effectively screened that the formation of bound states is inhibited. This kind of strongly interacting dense matter where the quarks, antiquarks and gluons behave collectively as free, deconfined, particles is called Quark-Gluon Plasma (QGP). Such a new phase of matter can be experimentally produced in ultrarelativistic heavy ion collisions (URHIC).

The phase transition between the confined and deconfined phases of QCD has been studied by the extensive ab initio calculations of lattice QCD. It has been shown that the QGP undoubtedly exists at sufficiently high energy densities. It has also been observed that chiral symmetry is restored at the same critical temperature as where the deconfinement phase transition takes place. For a purely gluonic (SU(3) gauge symmetric) system, for which the Equation of State (EoS) has been computed without approximations, the deconfinement phase transition is of the first order and the critical temperature is ${T_{c} \sim {150 MeV}}$. For the moment, dynamical quarks can only be included in certain approximations, and the order of the phase transition in full QCD is not yet known. The behavior of the QCD transition at non-zero baryochemical potential (at non-zero net-baryon number) is not yet known from first principles, either.

There are in principle two ways to achieve the QGP phase of matter: if ordinary nuclear matter is compressed to the extent that nucleons overlap, the quarks become deconfined and a cold QGP is formed. This situation may take place inside neutron stars. The experimentally relevant way is to prepare the QGP by ``heating'', i.e. by bringing energy into the system, out of which quark-antiquark pairs then form. Eventually, at sufficient densities and temperature, the quarks, antiquarks and gluons become deconfined and form a hot QGP. This is what happens in the URHIC: the energy for ``heating''(particle production) is provided by the collision energy, and the scattered quarks, antiquarks and gluons form the QGP. The bigger the nuclei are and the higher collision energy is, the more favorable are the conditions for the formation of the QGP.

The main goal of URHIC is to study the thermodynamics of strongly interacting matter and the QCD-phase transition in particular, i.e. to study condensed elementary particle matter. The field is quite interdisciplinary as it contains elements from particle physics, statistical field theory, particle kinetics, fluid dynamics and nuclear physics. The thermodynamics of strongly interacting matter has also a strong motivation in cosmology: according to the cosmological standard model our Universe underwent a QCD-transition from the QGP to the hadron gas a few microseconds after the Big Bang. In this sense in ultrarelativistic heavy ion collisions one is also studying ``Little Bang'' cosmology in a laboratory.

Experimentally the biggest challenge in colliding heavy nuclei (${A \sim 200)}$ at high center mass energies ${\sqrt{s} = 2...5500 AGeV}$ is to observe the QGP through specific probes sensitive to the presence of the QGP and to the occurrence of the QCD phase transition. In the theory of URHIC, calculations from truly first principles are practically impossible due to the complexity of the scattering dynamics and the non-perturbative features of the produced matter. What is needed from the theory of URHIC, however, is good phenomenology based on QCD. 

First hints of a collective strongly interacting system produced in URHIC were obtained from the fixed target experiments at the Super Proton Synchrotron (SPS) at CERN. The analysis of several observables from independent measurements also suggested that there are indications of the existense of the QGP in central lead-lead collisions at ${E_{beam}=158}$ ${AGeV}$ 
(cms-energy ${\sqrt{s}=17 AGeV}$). The actual properties of the QGP, in any case, clearly remain to be explored in the colliding-beam experiments.

The collider era of URHIC began in July 2000 at the Relativistic Heavy Ion Collider (RHIC) of the Brookhaven National Laboratory (BNL, NY, USA) with the Au-Au collisions at ${\sqrt{s}=56}$ and 130 AGeV. Since July 2001 the planned maximum cms-energy of RHIC, ${\sqrt{s}=200 AGeV}$, was reached. Also since Summer 2000 a truly impressive amount of high-quality data from several experiments at RHIC has been released, making the field more and more exciting. The experimental URHIC program will be completed by the measurements at the Large Ion Collider Experiment (ALICE) at the Large Hadron Collider (LHC) in CERN, which are still runing.

At much lower energies di-lepton data have also been taken the last years by the DSL collaboration at the BEVELAC, the CERES and the HELIOS-3 collaborations, the PHOENIX collaboration at RICH and aslo from HADES collaboration at SIS(GSI) with ${\sqrt{s} = 2 AGeV}$. In future (starting within 2012) the research in that direction will be continue using the replacement of the HADES collaboration which is the CBM (Combressed Baryonic Matter) at the same facilities at GSI in Darmstadt, Germany. A today status picture we can see at the Fig.1.1 in which depicted the QCD phase diagram. 

\begin{figure}[h]
\centering
\includegraphics[width=80mm, height=70mm]{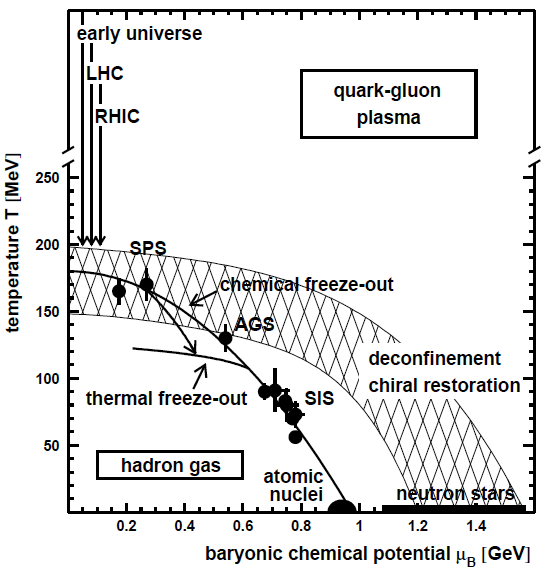}
\caption{QCD Phase diagram \cite{30}.}
\end{figure}

\subsection{Ultra Relativistic Heavy Ion Collisions and di-leptons}
We already know from many experiments, that as the nuclear matter heated and compressed hadrons occupy more and more of the available free space. Hadrons start to overlap each other and the initially confined quarks and gluons begin to percolate. This condition is the basis for models of the quark to hadron transition and has been essentially confirmed by numerical QCD lattice calculations at finite temperature. 

The above mentioned picture demonstrates that strongly interacting matter changes in energy and entropy density within a narrow temperature interval producing that we so-called Quark Gluon Plasma (QGP). At this conditions quarks which carrying an effective mass of a few hundred MeV in the confined phase lose their constituent mass which is proof of the restoration of chiral symmetry. Then almost massless left and right handed quarks decouple, leading to a degeneracy in states of opposite parity.

One of the most challenging tasks in experimental high energy physics is how to detect the hadron to quark-gluon transition in the heavy-ion collisions and most challenging is how can we isolate observable signals \cite{1},\cite{2}. Once of the most promising ways, because of their negligible final state interactions with the hadronic environment, are the di-leptons ${(e^{+}e^{-}}$ or ${\mu^{+}} {\mu^{-}})$ and photons. Are ideal probes for the high density or temperature areas which formed in the early stages of the collision and they sense in fact the entire spacetime history of the reaction \cite{3},\cite{4}. Because of their invariant pair mass, di-leptons have an advantage in the signal to background ratio, as compared to photons \cite{5}.

During the Ultra Relativistic Heavy Ion Collision the measured di-lepton spectra can separated into several moments-phases. At first, one of them takes place before the nuclear surfaces of the beam and the target actually touch. Then di-leptons are produced via a coherent Bremsstrahlung \cite{6} in the decelerating field of the approaching nuclei. Second within the first fm/c of the nuclear overlap, the excited hadronic system is far away from the thermal equilibrium and the corresponding  di-lepton radiation mostly consists of hard processes such as Drell-Yan annihilation in which can be seen at large invariant masses, ${m_{ll} \sim 3 GeV}$. Third is a rapid thermalization \cite{7} in which expected establishment of the QGP, which so-called partonic condition or phase, where di-lepton production proceeds via quark-antiquark annihilation.

At the incoming final stage upon expansion and cooling, the quark gluon plasma has converted into a hot hadron perfect liquid (after RHIC discovery, 2005) and then, di-leptons are radiated from ${\pi^{0,+,-}}$ and ${K^{+,-}}$ annihilation processes. Two-body annihilation processes are enhanced via the formation of vector mesons such as ${\rho}$, ${\omega}$ and ${\phi}$ and directly this mesons couple to ${l^{+}l^{-}}$ pairs. The most important thing in that process is that the invariant mass of the lepton pair, directly connected to the mass distribution of the vector meson at the moment of the decay. At the end, when the freeze-out stage is reached, the dominant sources are hadronic resonances and Dalitz decays from ${\pi^{0}}$, ${\eta}$ and ${\omega}$ mesons, all belong into the low-mass region, ${m_{ll } \le 1 GeV}$. A simple picture of the characteristic di-lepton sources in ultrarelativistic heavy-ion collisions (URHIC’s) as a function of invariant mass is given in Fig.1.2.

\begin{figure}[h]
\centering
\includegraphics[width=60mm, height=60mm]{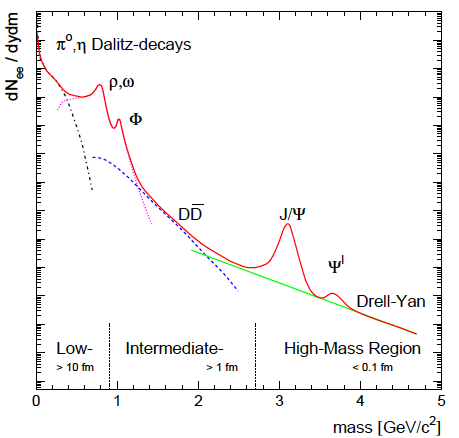}
\caption{Sources for di-lepton production as a function of invariant mass in URHIC's.\cite{30}}
\end{figure}

The last years a great activity has been made in various experiments and collaborations. One can start with CERES \cite{8, 9} and HELIOS-3 \cite{10} collaborations in which central A-A collisions exhibit a low-mass di-lepton production.
In Fig.1.3 we can see the invariant mass spectra of di-leptons from the CERES collaboration in 450 GeV proton-Beryllium and also proton-Gold at the same energy.

\begin{figure}[h]
\centering
\includegraphics[width=160mm, height=75mm]{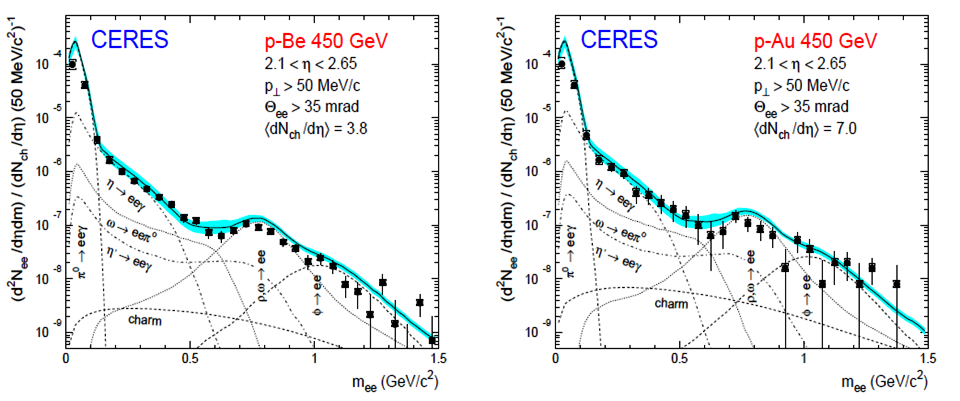}
\caption{Invariant mass spectra of di-leptons from CERES collaboration at 450 GeV in proton-Beryllium (left) and proton-Gold (right) at the same energy. \cite{8}}
\end{figure}

Also the same thing one can see in Fig.1.4 from CERES collaboration using Sulfur-Gold at 200 GeV producing  ${e^{+}e^{-}}$ di-lepton spectra and also from HELIOS-3, Sulfur to W target at the same energy producing  ${\mu^{+}\mu^{-}}$ di-lepton pairs.  
Finally we end with HADES collaboration, a di-lepton spectra from carbon-carbon (CC) and proton-proton (pp) systems in the Fig.1.5.\cite{11},\cite{12}

\begin{figure}[h]
\centering
\includegraphics[width=160mm, height=75mm]{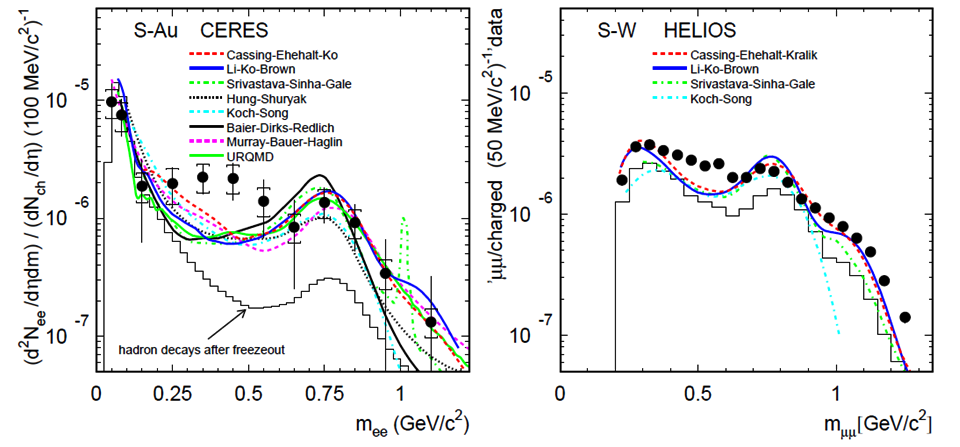}
\caption{Dileptom invariant mass spectra as measured in central collisions of 200 GeV from CERES \cite{8} using Sulfur-Gold (left) and from HELIOS-3 \cite{10} using Sulfur-W (right) at the same energy.}
\end{figure}

\begin{figure}[h]
\centering
\includegraphics[width=130mm, height=75mm]{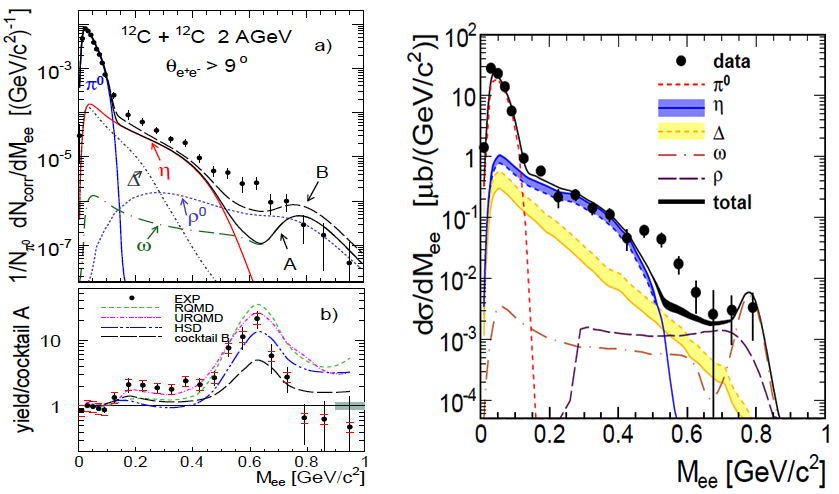}
\caption{Left: HADES Experiment ${^{12}C+^{12}C}$ at 2 AGev invariant mass distribution together with Monte Carlo simulations. Right: the same as before but now for proton proton at 2.2 GeV \cite{11},\cite{12}.}
\end{figure}

\section{QCD and Vector Mesons}
The standard Model (SM) tool for the strong interactions is the Quantum Chromodymanics (QCD). The QCD is a SU(3) gauge theory in which the quarks and the gluons are the elementary degrees of freedom. The QCD lagrangian is:

\begin{equation}
{{L}_{QCD}}=\bar{\psi }(i{{\gamma }^{\mu }}{{D}_{\mu }}-{{m}_{q}})\psi -\frac{1}{4}G_{\mu \nu }^{\alpha }G_{\alpha }^{\mu \nu }\label{eq:2.1}
\end{equation} 

where:
\begin{equation}
{{D}_{\mu }}={{\partial }_{\mu }}-ig\frac{{{\lambda }_{\alpha }}}{2}A_{\mu }^{\alpha }\label{eq:2.2}
\end{equation}
is the gauge covariant derivative which produces the connection between the fermionic color fields ${\psi}$ with ${N_{f}}$ different flavors and the gauge field $A_{\mu }^{\alpha }$, ${{\lambda }_{\alpha }}$ are the Gell-Mann SU(3) matrices.
Aslo becuse the QCD is a non-abelian theory the gluonic field-strength tensor written as:

\begin{equation}
G_{\mu \nu }^{\alpha }={{\partial }_{\mu }}A_{\nu }^{\alpha }-{{\partial }_{\nu }}A_{\mu }^{\alpha }+ig{{f}^{abc}}A_{\mu }^{b}A_{\nu }^{c}\label{eq:2.3}
\end{equation}
where $A_{\mu }^{\alpha }$ is the spin-1 gauge field for different colors $\alpha =1,2,3,4,5,6,7,8$.
In the QCD Lagrangian  eq.(\ref{eq:2.1}), the mass tern ${m_{q}}$ represents a diagonal matrix of the current quark masses of the up,down,strange which are the  'light flavors' (MeV) and the charm, top and bottom guarks which are the 'heavy flavors' (GeV).

To complete our view in the QCD theory we mast take into account the fact that due to quantum-loop effects the coupling constant ${\alpha_{s}}$ which connected with the factor g through the relation: 
${{\alpha }_{s}}=\frac{g}{4\pi }$ depends on the space distance. The coupling constant connected  via the four-momentum transfer Q of the strong process under the relation:

 \begin{equation}
{{\alpha }_{s}}(Q)=\frac{{{a}_{s}}(\Lambda )}{1+{{a}_{s}}(\Lambda )\frac{33-2{{N}_{f}}}{12\pi }\ln (\frac{{{Q}^{2}}}{{{\Lambda }^{2}}})}\label{eq:2.4}
\end{equation}
where ${\Lambda}$ is the coupling constant fixed by experiments with value around 200 MeV.  
As we can see from the Fig 2.1 as the Q increases  the  ${\alpha_{s}}$ decreases logarithmically and this is that we call 'asymptotic freedom'. in this area a pertrurbative theory can be applied.  For momentum scale less than 1GeV the picture is exactly the opposite. The  ${\alpha_{s}}$ increases as the Q decreases and the perturbation theory  is no longer applicable. This is the area where the hadrons formed and the confinement is strong.

\begin{figure}[h]
\centering
\includegraphics[width=90mm, height=75mm]{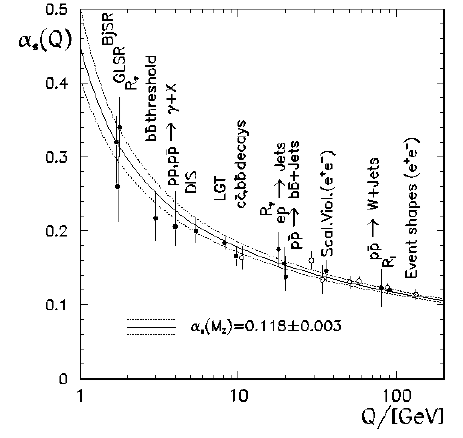}
\caption{The behavior of the  strong coupling  ${\alpha_{s}}$ as a function of the four-momentum Q transfer within various measurements \cite{13}.}
\end{figure}

\subsection{QCD Lagrangian Symmetries and the Axial Anomaly}
The "light flavor" hadrons, involving the up, down, and strange quarks  generally governed by chiral symmetry \cite{14} and its dynamical breaking in the physical vacuum. The confinement in this area plays a less role. Still today widely believed that pseudo-particles plays an important role in the hadronic correlators \cite{15},\cite{16} and is responsible for the chiral symmetry breaking and other non-perturbative phenomena. 

Apart from the fact that the QCD is invariand under local SU(3) transformations and a global U(1) symmetries, as for example a multiplication of the ${\psi}$  fields by a phase, lead us to a  baryon number conservation and then the QCD Lagrangian eq.(\ref{eq:2.1}) shows an additional symmetry for vanishing quark masses. 

In this approximate quark mass vanishing limit, which is well justified for up and down quarks,  the theory is invariant under global vector and pseudo-vector transformations in the SU(3) flavor space.

\begin{equation}
{\psi \to \exp \{-i\alpha _{V}^{i}\frac{{{\lambda }^{i}}}{2}\}\psi }\label{eq:2.5} 
\end{equation}
 \begin{equation}
{\psi \to \exp \{-i\alpha _{A}^{i}\frac{{{\lambda }^{i}}}{2}{{\gamma }^{5}}\}\psi }\label{eq:2.6}
\end{equation}

Looking to the corresponding currents from Noether's theorem we can write:

\begin{equation}
{j_{V,i}^{\mu }=\bar{\psi }{{\gamma }^{\mu }}\frac{{{\lambda }_{i}}}{2}\psi }\label{eq:2.7}
\end{equation}

\begin{equation}
{j_{A ,i}^{\mu }=\bar{\psi }{{\gamma }^{\mu }}{{\gamma }_{5}}\frac{{{\lambda }_{i}}}{2}\psi} \label{eq:2.8}
\end{equation}
 and the charges are:

\begin{equation}
{Q_{i}^{V}=\int{{{d}^{3}}x}{[{\psi }^{\dagger }}\frac{{{\lambda }_{i}}}{2}\psi ]}\label{eq:2.9}
\end{equation}

\begin{equation}
{Q_{i}^{A}=\int{{{d}^{3}}x}{[{\psi }^{\dagger }}\frac{{{\lambda }_{i}}}{2}{{\gamma }_{5}}\psi] }\label{eq:2.10}
\end{equation}
of course these charges commute with the QCD Hamiltonian: $[Q_{i}^{V,A},{{H}_{QCD}}]=0$.  
The quark fields can be decomposed to a right and left chiral parts and the QCD Lagrangian can be written as:

\begin{equation}
{{{L}_{QCD}}={{\bar{\psi }}_{L}}i{{\gamma }^{\mu }}{{D}_{\mu }}{{\psi }_{L}}+{{\bar{\psi }}_{R}}i{{\gamma }^{\mu }}{{D}_{\mu }}{{\psi }_{R}}-\frac{1}{4}G_{\mu \nu }^{\alpha }G_{\alpha }^{\mu \nu }-[{{\bar{\psi }}_{L}}{{m}_{q}}{{\psi }_{R}}+{{\bar{\psi }}_{R}}{{m}_{q}}{{\psi }_{L}}]}\label{eq:2.11}
\end{equation}
in which actually the right hand term mixes the quark masses dynamically. Also the transformations in eq.(\ref{eq:2.5}) and eq.(\ref{eq:2.6}) translated to:

\begin{equation}
{{{{\psi }_{L}}\to \exp \{-i\alpha _{L}^{i}\frac{{{\lambda }^{i}}}{2}{{\psi }_{L}}\}}   ,    \ \ \ \ \  {{{\psi }_{R}}\to {{\psi }_{R}}}}
\end{equation}

\begin{equation}
{{{{\psi }_{R}}\to \exp \{-i\alpha _{R}^{i}\frac{{{\lambda }^{i}}}{2}{{\psi }_{R}}\}}  ,    \ \ \ \ \  {{{\psi }_{L}}\to {{\psi }_L}}}
\end{equation}
 
At the limit of vanishing quark masses these transformations form a global $SU{{(3)}_{L}}\times SU{{(3)}_{R}}$  chiral symmetry for flavors. As a consequence the left and right quarks are not mixed any more dynamically and the handness then hold at the strong interactions.
If one wants to calculate the associated conserved charges then can have them in the form:

\begin{equation}
{Q_{i}^{L}=\int{{{d}^{3}}x[\psi _{L}^{\dagger }\frac{{{\lambda }_{i}}}{2}{{\psi }_{L}]}=\frac{1}{2}(Q_{i}^{V}-Q_{i}^{A})}}
\end{equation}

\begin{equation}
{Q_{i}^{R}=\int{{{d}^{3}}x[\psi _{R}^{\dagger }\frac{{{\lambda }_{i}}}{2}{{\psi }_{R}]}=\frac{1}{2}(Q_{i}^{V}+Q_{i}^{A})}}
\end{equation}

Most surprising is that taking in to account that the quark masses are equal to zero, the QCD Lagrangian then contains another symmetry. So  under axial global ${U_{A}(1)}$ transformations:

\begin{equation}
{\psi \to \exp \{-i\alpha {{\gamma }^{5}}\}\psi }
\end{equation}
the QCD Lagrangian eq.(\ref{eq:2.1}) is invariant. Although the Noether current is then: $j_{A,0}^{\mu }=\bar{\psi }{{\gamma }^{\mu }}{{\gamma }_{5}}\psi $.
Concerning a full quantum theory the divergence of  $j_{A,0}^{\mu }$ has an anomaly:

\begin{equation}
{{{\partial }_{\mu }}j_{A,0}^{\mu }=\frac{3}{8}{{\alpha }_{s}}G_{\mu \nu }^{\alpha }\tilde{G}_{\alpha }^{\mu \nu }}
\end{equation}
where:  $\tilde{G}_{\alpha }^{\mu \nu }={{\varepsilon }^{\mu \nu \kappa \beta }}G_{\kappa \beta }^{\alpha }$. This called the $U{{(1)}_{A}}$  axial anomaly. Concluding we can say that the QCD Lagrangian is symmetric only under the group of:

\begin{equation}
{SU{{(3)}_{L}}\times SU{{(3)}_{R}}\times U{{(1)}_{V}}} \label{eq:2.18}
\end{equation}
which implies a conservation of the vector and pseudo-vector currents.

\subsection{The Chiral Symmetry and its relation with the Vacuum Condensates}
It is well known that quarks and qluons condensate in the physical vacuum giving a non zero expectation values: $\left\langle \bar{\psi }\psi  \right\rangle$ \cite{17},\cite{18} and $\left\langle GG \right\rangle$ \cite{19} respectively. The gluon condensate can be considered as parameter connected with the non-perturbative regime effects. On the other hand a non zero quark condensate implies that the chiral symmetry is spontaneously broken. So the symmetry group of eq.(\ref{eq:2.18}) is broken down to:

\begin{equation}
{SU{{(3)}_{V}}\times  U{{(1)}_{V}}} \label{eq:2.19}
\end{equation}

According to this, the vector and the baryon current also conserved but on the other hand the QCD vacuum is not symmetric any more under pseudo-vector transformations (eq.(\ref{eq:2.6})). Chiral symmetry breaking in the domain of the light meson spectrum can be viewed by the two following ways:

\vspace{3 mm}
${i )}$  Eight almost massless Goldstone bosons (${\pi}$ ,K, ${\eta}$) they appeared and interact weakly at low energies.

${ii)}$ Parity doublets they do not exist. In particular scalar, pseudo-scalar and vector, pseudo-vector splitting for mesons. For the massless spin1/2 particles the helicity eigeinstates are also parity eigenstates. In the case where the chiral symmetry is unbroken we expect degenerate hadronic isospin multiplets with opposite parity. This is not of course observed in nature.
(in Fig.2.2 we present the experimental observed spectrum of the low mass mesons).  

\begin{figure}[h]
\centering
\includegraphics[width=90mm, height=75mm]{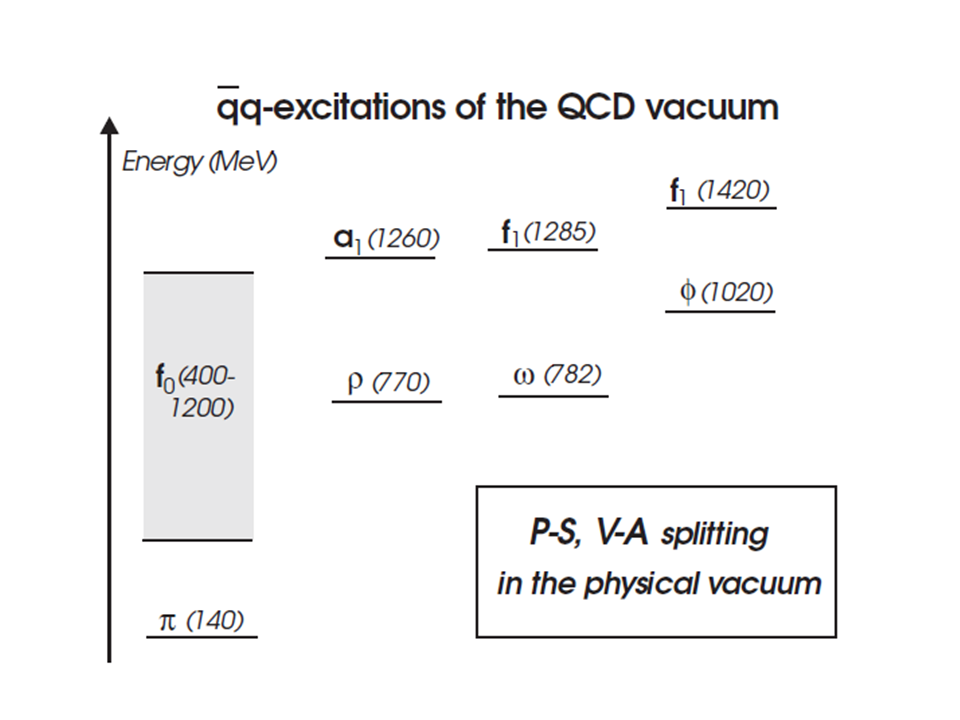}
\caption{The low-meson mass spectrum as observed experimentally \cite{30}.}
\end{figure}

Another implication of the chiral symmetry breaking is that the axial-current matrix element between the Goldstone boson and the vacuum is present. Examine the pions for example we can write:
\begin{equation}
{\left\langle  0 \right|j_{A,k}^{\mu }(x)\left| {{\pi }_{j}}(p) \right\rangle =-i{{\delta }_{ij}}{{f}_{\pi }}{{p}^{\mu }}\exp \{-ipx\}}
\end{equation}
where ${{f}_{\pi }}=93 MeV$  is the pion decay constant. This constant is the first order parameter which measures the strength of the symmetry breaking.  For the second order parameter we actually write the quark condensate:

\begin{equation}
{\left\langle \bar{\psi }\psi  \right\rangle \equiv \left\langle  0 \right|{{\bar{\psi }}_{L}}{{\psi }_{R}}+{{\bar{\psi }}_{R}}{{\psi }_{L}}\left| 0 \right\rangle }
\end{equation}
from this relation is obvious the mixing of right and left handed quarks in the QCD vacuum. The first order parameter (the pion decay constant) and the the quark condensate are related.  To know how, we must use the operator identity:
 \begin{equation}
{\left[ Q_{i}^{A},\left[ Q_{j}^{A},{{H}_{QCD}} \right] \right]={{\delta }_{ij}}\int{{{d}^{3}}x[\bar{\psi }}(x){{m}_{q}}\psi (x)]} \label{eq:2.22}
\end {equation}
Using the vacuum matrix element for different states we obtain the following sum rule: 

\begin{equation}
{\sum\limits_{n}{2{{E}_{n}}\left| \left\langle  n \right|Q_{i}^{A}{{\left. \left| 0 \right\rangle  \right|}^{2}}=-\left\langle {{m}_{u}}\bar{u}u+{{m}_{d}}\bar{d}d \right\rangle  \right.}=-2\bar{m}\left\langle \bar{q}q \right\rangle }\label{eq:2.23}
\end{equation}
 with ${\bar{m}}$ to be the average mass of up and down quarks. Taking in to account only pions we obtain the Gell-Mann-Oakes-Renner relation:

\begin{equation}
{m_{\pi }^{2}f_{\pi }^{2}=-2\bar{m}\left\langle \bar{q}q \right\rangle }\label{eq:2.24}
\end{equation} 
putting some values we can have for example, if ${\bar{m}} = 6 MeV$ then $\left\langle \bar{q}q \right\rangle =-{{(240MeV)}^{3}}=-1.8f{{m}^{-3}}$.
Because we interesting in for vector mesons a further useful parameter can be calculated also: 

\begin{equation}
{\left\langle j_{V,k}^{\mu }(x )j_{V,k}^{\mu }(0) \right\rangle -\left\langle j_{A,k}^{\mu }(x )j_{A,k}^{\mu }(0) \right\rangle }\label{eq:2.25}
\end{equation}

The eq.(\ref{eq:2.25}) is very important especially for the di-llepton production in heavy ion experiments. That because provides a connection between the spectral properties of the vector and pseudo-vector mesons and chiral symmetry breaking. 
Also for the in-medium modifications of the quark condensate will be important to take in account the matrix elements (for a any hadron) of the operator identity of eq.(\ref{eq:2.22}). That giving to us the sigma commutator for hadrons or the so-called '$\Sigma$-term'.

\begin{equation}
{{{\Sigma }_{hadron}}=\left\langle  h \right|[Q_{i}^{A},[Q_{j}^{A},{{H}_{QCD}}]\left| h \right\rangle =\left\langle  h \right|\bar{\psi }{{m}_{q}}\psi \left| h \right\rangle }\label{eq:2.26}
\end{equation}
 Making usu of Feynman-Hellmann theorem the $\Sigma$-term we be written in hadron and quark mass terms:
 
\begin{equation}
{{{\Sigma }_{hadron}}={{m}_{q}}\frac{\partial {{m}_{h}}}{\partial {{m}_{q}}}}\label{eq:2.27}
\end{equation}

The next step that we have to do is to examine the theory of the condensates in medium and especially when we consider low-density expansions.
 
\subsection{Condensates inside the medium}
When the temperature and the quark chemical potential are finite the quark condensate modified. Start using thermodynamics, the canonical partition function written as:

\begin{equation}
{Z (V,T,{{\mu }_{q}})=Tr[\exp \{-\frac{(\hat{H}-{{\mu }_{q}}\hat{N})}{T}\}]}\label{eq:2.28}
\end{equation}
using Hamiltonian and quark operators $\hat{H}$ and $\hat{N}$ respectively. For any operator ${\O}$ the thermal average written according to energy  ${E_{n}}$ and the exact eigenstates $\left| n \right\rangle $ as:

\begin{equation}
{\left\langle {\O} \right\rangle_{th} =\frac{1}{Z}\sum\limits_{n}{\left\langle n\left| {\O}\left| n \right. \right. \right\rangle }\exp \{-\frac{({{E}_{n}}-{{\mu }_{q}})}{T}\}}\label{eq:2.29}
\end{equation}
 
Applying the  eq.(\ref{eq:2.29}) to the QCD condensates we have for the quark and gluon condensates respectively:

\begin{equation}
{\left\langle \bar{\psi }\psi  \right\rangle_{th} =\frac{1}{Z}\sum\limits_{n}{\left\langle n\left| \bar{\psi }\psi \left| n \right. \right. \right\rangle }\exp \{-\frac{({{E}_{n}}-{{\mu }_{q}})}{T}\}}\label{eq:2.30}
\end{equation}

and 

\begin{equation}
{\left\langle GG \right\rangle_{th} =\frac{1}{Z}\sum\limits_{n}{\left\langle n\left| {{G}^{2}}\left| n \right. \right. \right\rangle }\exp \{-\frac{({{E}_{n}}-{{\mu }_{q}})}{T}\}}\label{eq:2.31}
\end{equation}

As next step we calculate the equation of states (EoS), taking the logarithm of the partition function ${Z}$. Then we obtain:

\begin{equation}
{\Omega =-\frac{{{T}^{2}}}{V}\ln Z}\label{eq:2.32}
\end{equation}
and for the energy density and the pressure we obtain respectively:

\begin{equation}
{\varepsilon =\frac{{{T}^{2}}}{V}{{(\frac{\partial \ln Z}{\partial T})}_{V,N}}+{{\mu }_{q}}\frac{N}{V}}\label{eq:2.33}
\end{equation}
and 
\begin{equation}
{p=T{{(\frac{\partial \ln Z}{\partial V})}_{T,N}}}\label{eq:2.34}
\end{equation}
when $V\to \infty $ and $\frac{N}{V}=const.$ (thermodynamic limit) then  the pressure and the free energy are equal.
Calculating the quark and gluon condensate from the EoS variables we get, first for quarks:

\begin{equation}
{\left\langle \bar{\psi }\psi  \right\rangle =\frac{\partial \Omega }{\partial {{m}_{q}}}=-\frac{\partial p}{\partial {{m}_{q}}}}\label{eq:2.35}
\end{equation}
 and using the thermal average of the trace of the energy-momentum tensor  $\left\langle T_{\mu }^{\mu } \right\rangle_{th} =\varepsilon -3p$,  for gluons:
\begin{equation}
{\left\langle GG \right\rangle _{th} =-\frac{8}{9}[(\varepsilon -3p)+{{m}_{q}}\frac{\partial p}{\partial {{m}_{q}}}]}\label{eq:2.36}
\end{equation}

At Ultrarelativistic Heavy Ion Collision experiments, as the temperature increases a population of thermally excited pions increases also. These excited pions represent the lightest hadrons. Taking into account a non interacting pion gas, the condensate then can be written as matrix element:  $\left\langle  \pi  \right|\bar{\psi }\psi \left| \pi  \right\rangle $. Using the ep.(\ref{eq:2.24}) and eq.(\ref{eq:2.27}) we obtain the following condensate ratio (in respect to vacuum):

\begin{equation}
{\frac{\left\langle \bar{\psi }\psi  \right\rangle_{th}}{\left\langle \bar{\psi }\psi  \right\rangle }\simeq 1-\frac{{{\Sigma }_{\pi }}\rho _{\pi }^{s }(T)}{f_{\pi }^{2}m_{\pi }^{2}}}\label{eq:2.37}
\end{equation}
where  ${\Sigma_{\pi}}$ and $\rho _{\pi }^{s}$ are the pion  '$\Sigma$-term' and scalar density in any given temperature.
The similar can be done for nucleons but now, ${\Sigma_{N}}$ and $\rho _{N}^{s}$ are the nucleon  '$\Sigma$-term' and scalar density for any given nucleon chemical potential. Usually ${{\mu }_{N}}\simeq 3{{\mu }_{q}}$.
\begin{equation}
{\frac{\left\langle \bar{\psi }\psi  \right\rangle_{th}}{\left\langle \bar{\psi }\psi  \right\rangle }\simeq 1-\frac{{{\Sigma }_{N}}\rho _{N}^{s}({{\mu }_{N}})}{f_{\pi }^{2}m_{\pi }^{2}}}\label{eq:2.38}
\end{equation}

Upon the creation of a hadron in the vacuum, the expectation value $\left\langle \bar{\psi }\psi  \right\rangle $ of the condensate, changes its value which is different of the vacuum.
The same development in the theory one can consider for the gluon condensates. In finite temperature the correction to he vacuum condensate should be given as before by a non-interacting pions. We know that the ${G^{2}}$ because is a chiral singlet its one pion matrix element disappears from the pion gas.  The things are different when we consider a finite baryon density. In our analysis we have to remind the formula for the nucleon mass:

\begin{equation}       
{{\left\langle  N \right|T_{\mu }^{\mu }\left| N \right\rangle} ={{m}_{N}}{{\bar{\psi }}_{N }}{{\psi }_{N }}}\label{eq:2.39}
\end{equation}
where 
\begin{equation} 
{T_{\mu }^{\mu }=-\frac{9}{8}{{G}^{2}}+\bar{\psi }{{m}_{q}}\psi}\label{eq:2.40}
\end{equation}
and
\begin{equation}   
{{{m}_{N}}=[-\frac{9}{8}\langle N|{{G}^{2}}\left| N \right\rangle +\langle N|\bar{\psi }{{m}_{q}}\psi \left| N \right\rangle ]}\label{eq:2.41}
\end{equation}
Here we note that: ${{G}^{2}}\equiv \frac{{{\alpha }_{s}}}{\pi }G_{\mu \nu }^{\alpha }G_{\alpha }^{\mu \nu }$.
In this relation eq.(\ref{eq:2.41}) according to eq.(\ref{eq:2.27}) the ${m_{q}}$ term is the nucleon  '$\Sigma$-term' and it is about 45 MeV.
Then taking all the above in account we extract the formula for the leading correction for the gluon condensates. As a function of a finite baryon density we can have:  
\begin{equation}
{\left\langle {{G}^{2}} \right\rangle_{th} -\left\langle {{G}^{2}} \right\rangle =-\frac{8}{9}m_{N(0)}\rho _{N}^{s}({{\mu }_{N}})}\label{eq:2.42}
\end{equation}
where ${m_{N(0)}}$ is the nucleon mass, comes from the matrix element $ {{G}^{2}}$.
To go further we assume that when baryon density vanishing the interaction inside the pion gas (studied through chiral perturbation theory) gives rise to a low temperature expansion of the condensate \cite{20},\cite{21},\cite{22}. One starting from the eq.(\ref{eq:2.35}) and eq.(\ref{eq:2.36}) and assuming that the quark mass is almost zero he can derive the relations for the pressure and the energy density.

\begin{equation}
{p=\frac{{{\pi }^{2}}}{90}(N_{f}^{2}-1){{T}^{4}}[1+N_{f}^{2}{{(\frac{{{T}^{2}}}{12f_{G}^{2}})}^{2}}\ln \frac{{{\Lambda }_{p}}}{T}]+O({{T}^{10}})}\label{eq:2.43}
\end{equation}

Making use of eq.(\ref{eq:2.35}), the quark mass derivative for mass ${m_{G}}$ (for the Goldstone boson) and the Gell-Mann-Oakes-Renner relation eq.(\ref{eq:2.24}) we can calculate again the ratio as:

\begin{equation}
{\frac{\left\langle \bar{\psi }\psi  \right\rangle_{th}}{\left\langle \bar{\psi }\psi  \right\rangle }=1+\frac{1}{f_{G}^{2}}\frac{\partial p}{\partial m_{G}^{2}}}\label{eq:2.44}
\end{equation}
and putting in this ratio the eq.(\ref{eq:2.43}) for the pressure we have: 

\begin{equation}
{\frac{\left\langle \bar{\psi }\psi  \right\rangle_{th}}{\left\langle \bar{\psi }\psi  \right\rangle }=1-\frac{(N_{f}^{2}-1)}{{{N}_{f}}}\frac{{{T}^{2}}}{12f_{G}^{2}}+\frac{(N_{f}^{2}-1)}{2{{N}_{f}}}{{(\frac{{{T}^{2}}}{12f_{G}^{2}})}^{2}}-{{N}_{f}}(N_{f}^{2}-1){{(\frac{{{T}^{2}}}{12f_{G}^{2}})}^{3}}\ln (\frac{{{\Lambda }_{q}}}{T})+O({{T}^{8}})}\label{eq:2.45}
\end{equation}
for ${N_{f}}=2$ the numerical value ${{\Lambda }_{q}}\simeq 470MeV$. 

Also the same thing again can be done for the gluons. To calculate a relation for low temperature about the gluon condensate (eq.(\ref{eq:2.36})) one uses the fact that:

 \begin{equation}
{\left\langle T_{\mu }^{\mu } \right\rangle_{th} =\varepsilon -3p={{T}^{5}}\frac{d}{dT}(\frac{p}{{{T}^{4}}})}\label{eq:2.46}
\end{equation}
Again making use of eq.(\ref{eq:2.43}) for the pressure we derive:
\begin{equation}
{\left\langle {{G}^{2}} \right\rangle_{th} -\left\langle {{G}^{2}} \right\rangle =-\frac{{{\pi }^{2}}}{3240}N_{f}^{2}(N_{f}^{2}-1)\frac{{{T}^{8}}}{f_{G}^{4}}[\frac{{{\Lambda }_{q}}}{T}-\frac{1}{4}]+...}\label{eq:2.47}
\end{equation}
Interpreting the previous relations one can extract some useful conclusions. When $(\varepsilon -3T)=0$ then $\left\langle T_{\mu }^{\mu } \right\rangle_{th} =0$ which means that we have a massless bosonic gas. This matches perfectly with the fact that the free gas of massless particles is scale invariant. Differences and changes at the gluon condensate arises of the interaction of Goldstone bosons, which are of course not scale invariant. Also the high order in temperature implies that the quark condensate melts faster than the gluon condensate.  

\subsection{Vector Mesons and Di-leptons}

At the begging of this review mentioned that one of the helpful ways to study the ultrarelativistic heavy ion collisions can be done via di-leptons production at the hot and dense medium. That because have found that these low mass di-leptons comes out almost unaffected carrying information about the process. These low mass di-leptons usually produced after quark and anti-quark or pion annihilation.  
For this process a useful relation can be calculated. This relation gives us the rate of di-leptons production at four-momentum from the heat bath at a given temperature \cite{23},\cite{24}:

\begin{equation}
{\frac{{{d}^{8}}{{N}_{{{l}^{-}}{{l}^{+}}}}}{{{d}^{4}}x{{d}^{4}}q}\equiv \frac{{{d}^{4}}R}{{{d}^{4}}q}={{L}^{\mu \nu }}(q){{W}_{\mu \nu }}(q)}\label{eq:2.48}
\end{equation}  
where ${{W}_{\mu \nu }}(q)$  and ${{L}_{\mu \nu }}(q)$ is the hadron and the leptons tensor respectively.
Taking in account the electromagnetic coupling constant:  $\alpha =\frac{1}{137}$ the leptons tensor can be written as:

\begin{equation}
{{{L}_{\mu \nu }}(q)=-\frac{{{\alpha }^{2}}}{6{{\pi }^{3}}{{M}^{2}}}({{g}_{\mu \nu }}-\frac{{{q}_{\mu }}{{q}_{\nu }}}{{{M}^{2}}})}\label{eq:2.49}
\end{equation}
In the following analysis we will focus on electron positron pairs. Inside the hadron tensor we can encode the effect of the medium. This shown at  following relation \cite{23},\cite{24}:

\begin{equation}
{{{W}_{\mu \nu }}(q)=\int{{{d}^{4}}x\exp \{iqx\}\left\langle \left\langle j_{m}^{EM}(x)j_{n}^{EM}(0) \right\rangle  \right\rangle }}\label{eq:2.50}
\end{equation}
Taking in account masses which are bellow the mass of the charm quark the electromagnetic current can be decomposed as:
\begin{equation}
{j_{\mu }^{EM}=\frac{2}{3}\bar{u}{{\gamma }_{\mu }}u-\frac{1}{3}\bar{d}{{\gamma }_{\mu }}d-\frac{1}{3}\bar{s}{{\gamma }_{\mu }}s}\label{eq:2.51}
\end{equation}
where ${{\gamma }_{\mu }}$ are the Dirac gamma matrices.

For each of the light vector mesons we can write an identification:  $j_{\mu }^{EM}=j_{\mu }^{\rho }+j_{\mu }^{\omega }+j_{\mu }^{\varphi }$ where:
\begin{equation}
{j_{\mu }^{\rho }=\frac{1}{2}(\bar{u}{{\gamma }_{\mu }}u-\bar{d}{{\gamma }_{\mu }}d) }\label{eq:2.52}
\end{equation}

\begin{equation}
{j_{\mu }^{\omega }=\frac{1}{6}(\bar{u}{{\gamma }_{\mu }}u+\bar{d}{{\gamma }_{\mu }}d) }\label{eq:2.53}
\end{equation}

\begin{equation}
{j_{\mu }^{\sigma }=-\frac{1}{3}(\bar{s}{{\gamma }_{\mu }}s) }\label{eq:2.54}
\end{equation}
are the currents for the ${\rho}$, ${\omega}$ and ${\phi}$ meson as a composition of their quarks.
In an a perfect plasma of quarks and gluons where the temperature is finite and almost zero chemical potential, the rate in eq.(\ref{eq:2.48}) can be  easily evaluated. Using low order perturbation theory and integrating over the 3-momentum of the di-lepton pair, one obtains:

\begin{equation}
{\frac{d{{R}^{q}}}{d{{M}^{2}}}={{R}^{q}}\frac{{{\alpha }^{2}}}{6{{\pi }^{2}}}MT[{{K}_{1}}(\frac{M}{T})]}\label{eq:2.55}
\end{equation}
 where ${{K}_{1}}(\frac{M}{T})$ is a Bessel function (modified) and ${{R}^{q}}={{N}_{c}}\sum\limits_{f}{e_{f}^{2}}=3(\frac{4}{9}+\frac{1}{9}+\frac{1}{9})=2$ for the different colors (${N_{c}}=3$).

At finite temperature we can derive an analogous relation (perfect gas resonances):

\begin{equation}
{\frac{d{{R}^{h}}}{d{{M}^{2}}}={{R}^{h}}(M)\frac{{{\alpha }^{2}}}{6{{\pi }^{2}}}MT[{{K}_{1}}(\frac{M}{T})]}\label{eq:2.56}
\end{equation}
where:  

\begin{equation}
{{{R}^{h}}=\frac{\sigma ({{e}^{+}}{{e}^{-}}\to h)}{\sigma ({{e}^{+}}{{e}^{-}}\to {{\mu }^{+}}{{\mu }^{-}})}}\label{eq:2.57}
\end{equation}
Using the rates at eq.(\ref{eq:2.55}) and eq.(\ref{eq:2.56}) and taking temperature close to the lattice results T=160 MeV for zero baryochemical potential, the previous rates are coincide around 1GeV and above. Also differences are well distinguished bellow 1GeV due to  ${\rho}$, ${\omega}$ and ${\phi}$ resonances as depicted at Fig.2.3.

\begin{figure}[h]
\centering
\includegraphics[width=60mm, height=80mm]{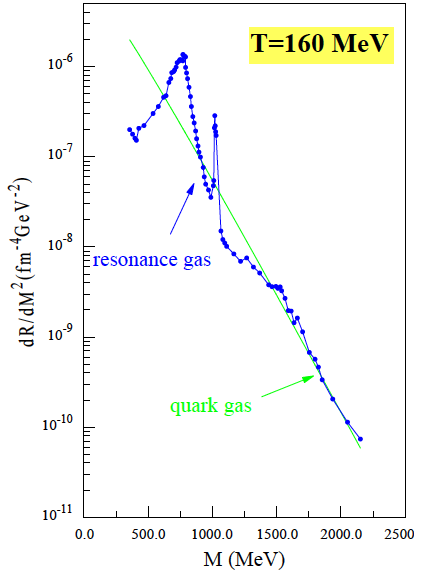}
\caption{Three-momentum integrated di-lepton production rate using  first perturbative quark-antiquark prediction (eq.(\ref{eq:2.55})) and second resonance gas approximation (eq.(\ref{eq:2.56})).\cite {30}}
\end{figure}

\subsection{The Sum Rules of QCD}
In general the QCD sum rules, invented by Shifman, Vainshtein and Zakharov and is a very interesting approach for theoretical study of the structure of the hadrons. Is applicable to the simplest static hadronic characteristics, like mass, leptonic widths etc., they also used to calculate much more complicated things like hadronic wave functions and form factors.
\subsubsection{In-Vacuum}
The sum rules of QCD \cite{25} is an attempt to interpret the current-current correlation functions in terms of QCD. The way that one can do this is by relating the hadron spectrum with the non-perturbative QCD structure of the vacuum.
More practically an easy way to achieve this, is to separate the analysis in to scales, one for sort and one for long distances. Tool for that is the "Operator Product Expansions"(OPE) which connects the time-like light quark currents, to big space-like momenta:  

\begin{equation}
{{{Q}^{2}}\equiv -{{q}^{2}}>0}\label{eq:2.58}
\end{equation}
A gauge invariant local operators ${\O}$ (actually a series of them) related with this output as:

\begin{equation}
 {-i\int{{{d}^{4}}x}[\exp \{iq\cdot x\}]T{{j}^{\mu }}(x){{j}^{\nu }}(0)=-({{g}^{\mu \nu }}-\frac{{{q}^{\mu }}{{q}^{\nu }}}{{{q}^{2}}})\sum\limits_{n}{{{c}_{n}}({{Q}^{2}},{{\Lambda }^{2}})}{{\O}_{n}}({{\Lambda }^{2}})}\label{eq:2.59}
\end{equation}
where ${\Lambda}$ is the scale of the renormalization and ${{c}_{n}}$  are the Wilson coefficients which basically contain the physics at short distances.  The long distance physics of QCD encapsulated inside the ${Q^{2}}$ independent ${\O_{n}}$ operators and manifests with the appearance of quark an gluon condensates. The above mentioned operators are multi-dimensional (D) but most of them especially in high orders are suppressed.  
The vacuum matrix elements of eq.(\ref{eq:2.59}) together with the following dispersion relation:

\begin{equation}
{{{\Pi }_{0}}({{Q}^{2}}=-{{q}^{2}})={{\Pi }_{0}}(0)+{{Q}^{2}}\int\limits_{0}^{\infty }{ds\frac{{{\rho }_{0}}(s)}{{{Q}^{2}}+s}}}\label{eq:2.60} 
\end{equation}
where  
\begin{equation}
{{{\rho }_{0}}(s)=-(\frac{1}{\pi s}){Im}({{\Pi }_{0}}(s))}\label{eq:2.61} 
\end{equation}
is the vacuum spectral function of the current-current correlator for electromagnetic current ${{\Pi}_{0}(0)=0}$ (massless photon).
Next goal is to calculate the Wilson coefficients. Using the left hand side of the eq.(\ref{eq:2.60}) on obtains a series of power expansion in ${{Q}^{2}}$ for dimension D:

\begin{equation}
{\frac{12\pi }{{{Q}^{2}}}{{\Pi }_{0}}({{Q}^{2}})=\frac{D}{\pi }[-{{c}_{0}}\ln (\frac{{{Q}^{2}}}{{{\Lambda }^{2}}})+\frac{{{c}_{1}}}{{{Q}^{2}}}+\frac{{{c}_{2}}}{{{Q}^{4}}}+\frac{{{c}_{3}}}{{{Q}^{6}}}+...]}\label{eq:2.62} 
\end{equation}
For the ${\rho}$ meson \cite{26} we obtain:

\begin{equation}
{{ c_{0}^{\rho }=1+\frac{{{\alpha }_{s}}}{\pi }}}\label{eq:2.63} 
\end{equation}
\begin{equation}
 {{ c_{1}^{\rho }=-3(m_{u}^{2}+m_{d}^{2}) }}\label{eq:2.64} 
\end{equation}

\begin{equation}
{c_{2}^{\rho }=\frac{{{\pi }^{2}}}{3}\left\langle {{G}^{2}} \right\rangle +4{{\pi }^{2}}\left\langle {{m}_{u}}\bar{u}u+{{m}_{d}}\bar{d}d \right\rangle}\label{eq:2.65} 
\end{equation}

\begin{equation}
{c_{3}^{\rho }\propto {{\alpha }_{s}}\left\langle {{(\bar{q}q)}^{2}} \right\rangle }\label{eq:2.66}
\end{equation}
 One useful tool to continue our review is to know about the Borel mass and the Borel transformation:

\begin{equation}
{{{\hat{L}}_{M}}=\lim \frac{1}{(n-1)!}{{({{Q}^{2}})}^{n}}{{(-\frac{d}{d{{Q}^{2}}})}^{n}}}\label{eq:2.67}
\end{equation}
where ${M}$ is the Borel mass and the limits are ${{Q}^{2}}\to \infty $ , $n\to \infty $ and $\frac{{{Q}^{2}}}{n}={{M}^{2}}$. Using this to eq.(\ref{eq:2.60}) from both sides we obtain:

\begin{equation}
{\frac{1}{\pi {{M}^{2}}}\int{ds[{{\rho }_{o,V}}(s)]\exp \{-\frac{s}{{{M}^{2}}}\}=\frac{{{d}_{V}}}{12{{\pi }^{2}}}[{{c}_{0}}+\frac{{{c}_{1}}}{{{M}^{2}}}+\frac{{{c}_{2}}}{{{M}^{4}}}+\frac{{{c}_{3}}}{2{{M}^{6}}}+....]}}\label{eq:2.68}
\end{equation}
where ${d_{V}}$ is the continuum strength for each vector meson (light in our case), ${\rho_{0,V}(s)}$ is the vacuum spectral function (density) for vectors and the Borel mass limit is around 1 GeV.

One hypothesis that can we make here is that the vacuum spectral density follows a delta function parametrization of the resonance part,  supported by a consecutive step function (Fig.2.4): 

\begin{equation}
{{{\rho }_{0}}(s)=\frac{{{z}_{V}}}{12{{\pi }^{2}}}\delta (s-m_{V}^{2})+\frac{{{d}_{V}}}{12{{\pi }^{2}}}(1+\frac{{{\alpha }_{s}}}{\pi })\Theta (s-{{s}_{V}})}\label{eq:2.69}
\end{equation}
${z_{V}}$ stands for the pole strength of the vector meson, ${m_{V}}$ is the vector mass and ${d_{V}}$ denote the consecutive threshold.

\begin{figure}[h]
\centering
\includegraphics[width=100mm, height=60mm]{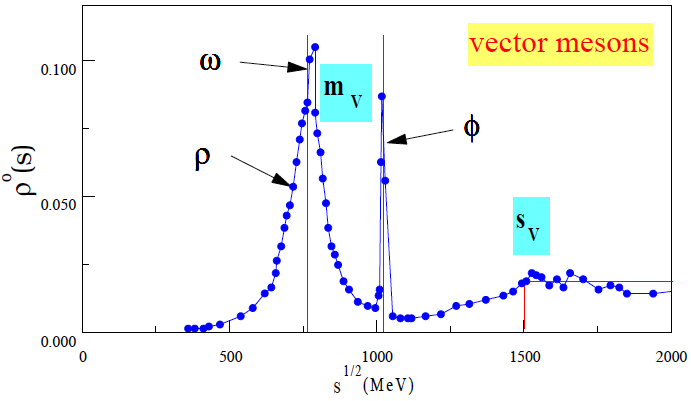}
\caption{Parametrization of the strength function in QCD sum rules\cite{30}.}
\end{figure}

\subsubsection{In-Medium}
The QCD sum rules are different in-medium than in vacuum. In a hot hadronic medium the vector current-current correlation function (for a hadron gas) given by:

\begin{equation}
{{\Pi^{\mu\nu}_{A}}(q)=-i\int d^{4}x[exp\left\{  iq\cdot x\right\}]\left\langle j^{\mu}_{A}(x) j^{\nu}_{A}(0) \right\rangle_{th}}\label{eq:2.70}
\end{equation}
The thermal medium breaks explicitly Lorentz invariance. The structure tensor becomes more complicated and gives a separate dependence on $q_{0}$ (energy) and the 3-momentum $\overrightarrow{q}$. Then a longitudinal and transverse components appearing:

\begin{equation}
{{{\Pi^{\mu\nu}_{V,A}}(q)={{\Pi^{L}_{V,A}}(q_{0},\overrightarrow{q})P^{\mu\nu}_{L}+{\Pi^{T}_{V,A}}(q_{0},\overrightarrow{q})P^{\mu\nu}_{T}}}}\label{eq:2.71}
\end{equation}
${P_{L}}$ and ${P_{T}}$ are the longitudinal and transverse projection operators respectively: 

\begin{equation}
{P^{\mu\nu}_{L}=\frac{q^{\mu}q^{\nu}}{M^{2}}-g^{\mu\nu}-P^{\mu\nu}_{T}}\label{eq:2.72}
\end{equation}

$$
P^{\mu\nu}_{T} = \left\{ \begin{array}{lc}
 0 &\mbox{ if  $\mu=0${ or $\nu=0$}}\\
  \delta^{ij}-\frac{q^{i}q^{j}}{q^{2}} &\mbox{ for  $\mu$, $\nu$}{\in\left\{1,2,3\right\}}  
       \end{array} \right.
$$
where here the space-like componets have replaced from $i$ and $j$ insted of ${\mu}$ and ${\nu}$ respectively.
Assuming the case where $\vec{q}=0$ for which  $\Pi _{V}^{L,T}({{q}_{0}})$ of eq.(\ref{eq:2.71}) concur. In almost one to one analogy with eq.(\ref{eq:2.68}) the in-medium sum rules at T=0 given by:

\begin{equation}
{\frac{1}{\pi {{M}^{2}}}[{{\Pi }_{V}}(0)+\int{dq_{0}^{2}{{\rho }_{V}}({{q}_{0}})\exp \{-\frac{q_{0}^{2}}{{{M}^{2}}}\}}]=\frac{{{d}_{V}}}{12{{\pi }^{2}}}[{{c}_{0}}+\frac{{{c}_{1}}}{{{M}^{2}}}+\frac{{{c}_{2}}(\rho )}{{{M}^{4}}}+\frac{{{c}_{3}}(\rho )}{2{{M}^{6}}}+...]}\label{eq:2.73}
\end{equation}
in which ${{\rho }_{V}}({{q}_{0}})$ is the in-medium vector spectral function:

\begin{equation}
{{{\rho }_{V,A}}({{q}_{0}})=-\frac{1}{q_{0}^{2}\pi }\Pi _{V,A}^{L}({{q}_{0}},0)}\label{eq:2.74}
\end{equation}

Examine more carefully the right hand side of the eq.(\ref{eq:2.73}) we can see the entrance of the medium through the density dependent Wilson coefficients $c_{2}(\rho)$ and  $c_{3}(\rho)$. These coefficients are determined by the guark and gluon density dependent condensates:

\begin{equation}
{{{\left\langle \bar{q}q \right\rangle }_{th}}=\left\langle \bar{q}q \right\rangle -\frac{{{\Sigma }_{N}}}{2\bar{m}}\rho }\label{eq:2.75}
\end{equation}

\begin{equation}
{{{{\left\langle {{G}^{2}} \right\rangle }_{th}}=\left\langle {{G}^{2}} \right\rangle -\frac{8}{9}{{m}_{N(0)}}\rho }}\label{eq:2.76}
\end{equation}

In \cite{27} is pointed out that an extra contribution to Wilson coefficients probably coming from new condensates which mixing the quark and gluon fields. These appearing through the covariant derivative $D_{\mu}$ (eq.(\ref{eq:2.2})). The matrix elements are analogous to quark and anti-quark distribution function at the nucleon:  

\begin{equation}
{A_{n}^{q}=2\int\limits_{0}^{1}{dx({{x}^{n}}[q(x)+\bar{q}(x)])}}\label{eq:2.77}
\end{equation}
According to \cite{27} for the $\rho$  meson we can write:

\begin{equation}
{c_{2}^{\rho }(\rho )\simeq c_{2}^{\rho }(0)-(\frac{8{{\pi }^{2}}}{27}{{m}_{N(0)}}-2{{\pi }^{2}}A_{1}^{u+d}{{m}_{n}})\rho }\label{eq:2.78}
\end{equation}
and 
\begin{equation}
{c_{3}^{\rho }(\rho )\simeq c_{3}^{\rho }(0)-(\frac{896}{81}k {{\pi }^{3}}{{\alpha }_{s}}\frac{{{\Sigma }_{N}}}{{\bar{m}}}\left\langle \bar{q}q \right\rangle -\frac{10}{3}{{\pi }^{2}}A_{3}^{u+d}m_{n}^{3})\rho }\label{eq:2.79}
\end{equation}
and according to \cite{28}, the nucleon mass at the chiral limit calculated to be ${{m}_{N(0)}}\simeq 750MeV$. Furthermore if one wants to see the Wilson coefficients also for the other meson, the \cite{26} is a good guide. 
Finally to complete all the necessary relations for the in-medium sum rules, the counterpart spectral function of the eq.(\ref{eq:2.69}) for vacuum now can be written as: 

\begin{equation}
{{{\rho }_{V}}({{q}_{0}})=\frac{1}{12{{\pi }^{2}}}[z_{V}^{*}\delta (q_{0}^{2}-m_{V}^{*2})+{{d}_{V}}(1+\frac{{{\alpha }_{s}}}{\pi })\Theta (q_{0}^{2}-s_{V}^{*})]}\label{eq:2.80}
\end{equation}
In \cite{27} one can find the medium dependence for the $\rho$ or  $\omega$ meson masses through the relation:

\begin{equation}
{\frac{m_{\rho ,\omega }^{*}}{{{m}_{\rho ,\omega }}}=1-(0.18\pm 0.06)\frac{\rho }{{{\rho }_{0}}}}\label{eq:2.81}
\end{equation}
One very profound implication-conclusion which comes out from eq.(\ref{eq:2.81}) is that, the masses decrease as the density increases. This is an indication of the Brown-Rho mass dropping scenario \cite{29}. 

From experiments vector meson mass modifications in-medium was reported by CERES experiment at SPS-CERN \cite{31},\cite{32}. Significant enhancements were observed in the range of 0.2 - 0.8 ${GeV/c^{2}}$ in the spectra, which were not observed before. In the literature, two representative scenarios, the mass dropping (Brown-Rho scaling)\cite{29} and the width broadening (by Rapp and Wambach)\cite{30} of the ${\rho}$ meson are compared with the data using the same space-time evolution of the fireball by Rapp. However, the data can be explained by both scenarios, namely no definite interpretation can be selected. The next experiment performed at SPS was the NA60. They measured ${\mu^{+}\mu^{-}}$ pairs with a good mass resolution and very large statistics in comparison with CERES. The enhancement is observed again below the ${\omega}$ meson peak\cite{33}. They claimed that the enhancement was explained by the width broadening of ${\rho}$ meson based on the in-medium many body calculation and the mass-dropping scenario (Brown-Rho scaling) was excluded. The data are reviewed in\cite{34}. According to \cite{35}, the in-medium ${\rho}$ width is 350 ${MeV/c^{2}}$, which is the average value over the space-time evolution of the fireball. CERES also claimed that their new data obtained in 2002 for 158 AGeV Pb+Au, are also explained by such a width broadening of the ${\rho}$ meson\cite{36}. The broadening scenario has gained predominance, however the objections from the other part still exist\cite{37},\cite{38}. PHENIX experiment also reported the observation of the enhancement below the ${\omega}$ meson peak\cite{39} measuring ${e^{+}e^{-}}$ spectrum in Au+Au collisions at 200 GeV in center mass frame. 
 
\section{Conclusions}
This article had as a goal to present the appropriate minimal theoretical background for the study of the QCD and Vector Mesons in a di-leptonic Quark Gluon Plasma. The hadron properties and the investigation of them in a hot and dense environment such as produced at the Ultra Relativistic Heavy Ion Collisions experiments today is one of he most interesting parts in theoretical and experimental high energy physics. This research is directly related to the hunting for the QCD phase transition. The study of the di-leptons which are produced in such collisions is a very useful tool. The final states of di-leptons are mediated by electromagnetic currents and the connection to the vector mesons working as a useful tool for acquiring information for the in-medium effects. Examining the low mass region of the vector mesons such as $\rho$, $\omega$ and $\phi$ we notice that the $\rho$ meson plays an important and dominant role because has a very short lifetime (the shortest) and the largest di-lepton decay width. Today is widely believed that the non-pertrurbative feature of low-energy strong interactions which is the spontaneously break of the chiral symmetry, is not only responsible to build the constituent quark masses but also for the existence of the low-energy hadrons spectrum. Hence the research for the chiral symmetry restoration in that hot and dense matter is directly related to the in-medium changes of the properties of the light hadrons. At the chiral symmetry restoration phase the chiral partners are becoming identical. This notion exist at the Weinberg sum rules in which at the chiral limit the pion decay constant $f_{\pi}$ relates the energy-integrated difference between vector and pseudo-vector correlators. 
Looking at the QCD sum rules we saw that they provide a relation between the vacuum correlators and the condensates, giving us a link between physical observed hadrons and the QCD vacuum structure. Is clear that when these rules applied at the hadronic medium, they have a limited power to predict the in-medium spectral distributions and they provide a band of allowed combinations of small-large masses and corresponding small-large decay widths for the vector mesons.

The theoretical and experimental efforts towards to understanding the nature and the underlying physics behind of the Ultra Relativistic Heavy ion Collisions continuing. Experiments as RHIC in United States, ALICE and  NA61 at LHC-CERN in Geneva, HADES and CBM in Darmstadt-Germany  and many others bigger or smaller facilities in different places still hunting for the remaining (after Higgs discovery this year) "Holy Grail" in heavy ion collisions physics, the observation of the deconfinement phase of a hadron.     

\section{Acknowledgments}
The author would like to thank the Pr. R. Rapp and Pr. J. Wambach for their useful papers which mostly have been used as a guide to this short review article.

\end{document}